\documentclass[fleqn,usenatbib]{mnras}
\usepackage{newtxtext,newtxmath}
\usepackage[T1]{fontenc}

\DeclareRobustCommand{\VAN}[3]{#2}
\let\VANthebibliography\thebibliography
\def\thebibliography{\DeclareRobustCommand{\VAN}[3]{##3}\VANthebibliography}

\usepackage{graphicx}	
\usepackage{amsmath}	
\usepackage{xcolor}

\title[Detecting post-merger signals in noisy GW detector data]{The effect of noise artefacts on gravitational-wave searches for neutron star post-merger remnants}

\author[F. H. Panther et al.]{
Fiona H. Panther,$^{1,2}$\thanks{E-mail: fiona.panther@uwa.edu.au}
Paul D. Lasky,$^{2,3}$
\\
$^{1}$Department of Physics, University of Western Australia, Crawley WA 6009, Australia\\
$^{2}$OzGrav: The ARC Centre of Excellence for Gravitational-wave Discovery\\
$^{3}$School of Physics and Astronomy, Monash University, VIC 3800
}

\date{Accepted XXX. Received YYY; in original form ZZZ}

\pubyear{2022}

\begin{document}
\label{firstpage}
\pagerange{\pageref{firstpage}--\pageref{lastpage}}
\maketitle

\begin{abstract}
Gravitational waves from binary neutron star post-merger remnants have the potential to uncover the physics of the hot nuclear equation of state. These gravitational-wave signals are high frequency ($\sim$ kHz) and short lived ($\mathcal{O}(10\,\mathrm{ms})$), which introduces potential problems for data-analysis algorithms due to the presence of non-stationary and non-Gaussian noise artefacts in gravitational-wave observatories. We quantify the degree to which these noise features in LIGO data may affect our confidence in identifying post-merger gravitational-wave signals. We show that the combination of vetoing data with non-stationary glitches and the application of the Allen $\chi^2$ veto (usually reserved for long-lived lower-frequency gravitational-wave signals), allows one to confidently detect post-merger signals with signal-to-noise ratio $\rho\gtrsim8$. We discuss the need to incorporate the data-quality checks and vetos into realistic post-merger gravitational-wave searches, and describe how one can incorporate them to calculate realistic false-alarm and false-dismissal rates.
\end{abstract}




\section{Introduction}
When two neutron stars merge, they may promptly collapse to form a black hole, or form a differentially-rotating neutron star that may be either stable or quasi-stable \cite[see][for a review]{Sarin2021_rev}.
During the first $\lesssim 100\,\mathrm{ms}$ following merger, numerical-relativity simulations indicate that gravitational waves are emitted by such a stable or quasi-stable post-merger remnant. These gravitational waves can be used as a sensitive probe of the equation of state of nuclear matter~\citep[e.g.,][]{Bauswein2012_NSproperties,Bauswein2012_eosdependence,Read2013,Takami2014}. 
To date, two binary neutron star mergers have been observed by the LIGO-Virgo-KAGRA \citep[LVK;][]{adv_ligo_2015, AdvancedVirgo, 2020_Kagra} collaboration: GW170817~\citep{GW170817} and GW190425~\citep{abbott21_190425}.
These events have given some of the strongest constraints to date on the cold nuclear equation of state based on measurements of the neutron stars' tidal deformability and mass during the final moments of the inspiral~\citep{GW170817, GW170817_props, abbott21_190425}.
However, gravitational waves from post-merger remnants have not yet been detected~\citep{abbott17_postmerger,abbott19_postmerger,abbott21_190425}.

The properties of the post-merger waveform depend on the internal structure and composition of the newborn neutron star.
For example, numerical-relativity simulations indicate that the dominant oscillation frequency of the quadrupolar f-mode is a sensitive probe of the compactness and tidal deformability of the remnant~\citep{Bauswein2012_eosdependence, Bauswein2012_NSproperties, Takami2014, Hotokezaka2013, Bauswein2019}.
Identifying the post-merger gravitational waves and measuring this frequency---expected to occur between $\sim 2-4\,\mathrm{kHz}$ depending on the exact equation of state---would enable measurements of the hot nuclear equation of state, providing complementary information to that gleaned about the cold equation of state measured during the inspiral.
For example, this has the potential to uncover temperature-dependent phase transitions that may occur in ultra-dense, hot nuclear matter~\citep[e.g.,][]{Bauswein2019_phasetransition}.

Previous works have developed a variety of phenomenological models that are used to perform parameter estimation for post-merger gravitational waves.
These models are based on quasi-universal relations between the frequency components of the waveform and physical parameters: neutron star radius, tidal deformability, and compactness.
These models have the capability of capturing the key features of numerical-relativity simulations while also being computationally tractable~\citep{Bauswein2016, Easter2020, Clark2016, Whittaker2022, Breschi2022}.
Many of these works have indicated that a robust detection could be made for a post-merger signal-to-noise ratio $\rho\gtrsim8$~\citep[e.g.,][]{Easter2020}\footnote{We note that~\citet{Clark2016} define a signal-to-noise threshold of $\rho=5$ without exploring the consequences of this choice, while \citet{Chatziioannou2017} show post-merger signals can be analysed as low as $\rho\gtrsim5$, however they use zero-noise realisations to understand parameter estimation rather than detection.}.
However, all of these works focus on signals injected into idealised Gaussian noise realisations, ignoring the effects of non-stationary and non-Gaussian noise artefacts (often called \textit{glitches} in LVK parlance).
Many of these glitches are short-lived and have a relatively high frequency component \citep[$\gtrsim500 {\rm Hz}$, ][]{Davis2021}. They may therefore mimic that of a post-merger gravitational-wave signal.

In this work, we investigate our ability to distinguish transient noise artefacts from gravitational-wave signals of post-merger remnants. We therefore determine how loud typical signals must be to reach a given false-alarm rate threshold.
In Section \ref{section:method} we describe our method to identify signal-like triggers using representative post-merger waveforms together with matched filtering. We discuss the results of filtering simulated Gaussian and real detector noise using our algorithm in Section \ref{section:results}, and calculate the false-alarm rate as a function of signal-to-noise ratio. We also discuss various mitigation strategies that could be applied to improve the sensitivity of such a search, as well as the potential limitations these may introduce in a realistic scenario.

\section{Methods}{\label{section:method}}
The matched filter is the optimal algorithm for detecting a signal of known form in Gaussian noise, commonly employed in the detection of gravitational waves \citep[see][for a review]{LIGOdata_rev}.
We use this technique as it affords a greater sensitivity and ability to reject glitches than typical unmodelled short-duration burst searches.
To search for a known post-merger gravitational wave signal $h(t, \vec{\theta})$ parameterised by a vector of observables $\vec{\theta}$ in detector data $d(t)$, we calculate the time-dependent complex matched-filter output between the template and data at time $t_0$ denoting the end time of the filter. This is given by 
$$
z(t_0) = 4\int_{0}^\infty \frac{\tilde{d}(f)\tilde{h}^*(f, \vec{\theta})}{S_n(f)}\exp{(2\pi i f t_0)}df,
$$
where tilde denotes a Fourier transform, $^*$ the complex conjugate, and $S_n(f)$ is the one-sided noise power-spectral density (PSD).
We search the detector strain data by varying the end time of the template $t_0$.
This filter output is used to create a detection statistic---the signal-to-noise ratio---by normalizing each template by the factor
$$
\sigma^2 = 4\int_{0}^\infty \frac{|\tilde{h}(f, \vec{\theta})|^2}{S_n(f)}df
$$
which accounts for the sensitivity of the instrument to a given signal. The signal-to-noise ratio, denoted $\rho$, is then
$$
\rho(t) = \frac{|z(t)|}{\sigma}.
$$
In this work, we use the \texttt{pycbc} software package \citep{PyCBC} to implement our detection workflow, which computes the complex two-phase filter output. We then maximise over the unknown phase of the signal by using the absolute value of the filter output to compute $\rho$.
We truncate the signal-to-noise ratio computation at $f_\mathrm{high} = 4096\,\mathrm{Hz}$ to account for the finite sampling rate of the data.

We use a post-merger waveform filter that captures the phenomenology of numerical-relativity simulations. 
We use a waveform comprised of three exponentially-damped sinusoids \citep[as described in][]{Bauswein2016}.
This waveform is also equivalent to that described in~\cite{Easter2020} with the quadratic frequency drift term set to zero. This allows for an analytically-tractable frequency-domain expression for the `plus'-polarisation of the strain: 
$$
\tilde{h}_\mathrm{+}(f,\vec{\theta}) = H\sum_{j=0}^2 \frac{\xi_j}{2}\bigg[\frac{\exp(i\psi_j)}{T_j^{-1} - 2\pi i(f_j - f)} + \frac{\exp(-i\psi_j)}{T_j^{-1} + 2\pi i(f_j + f)}\bigg]
$$
where the waveform is parameterised by $\vec{\theta}=[H,\xi,T,f, \psi]$ described in Table \ref{tab:params}.
Note that $H$ is an arbitrary normalisation that appears in both the numerator and denominator of the signal-to-noise ratio calculation, implying it's specific value is unimportant.
The cross polarisation $h_\mathrm{\times}$ component differs from the $h_\mathrm{+}$ component by a phase shift of $\pi/2$, and the contribution of the $h_\mathrm{\times}$ component to the total signal-to-noise ratio is encoded in the complex component of the filter output $z(t)$.

\begin{figure}
	 		\includegraphics[width=\columnwidth]{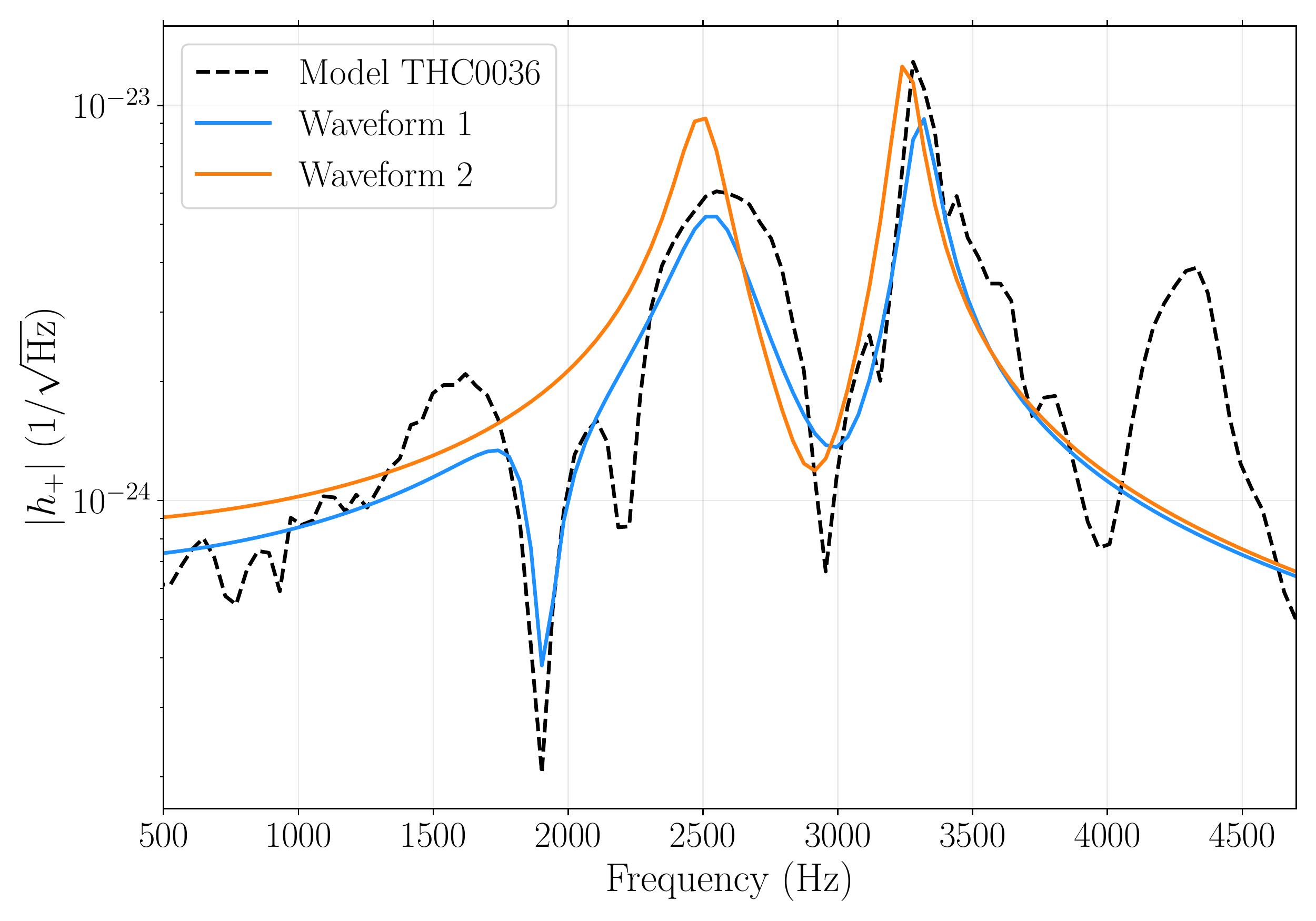}
	 		\caption{\label{fig:waveform}Numerical-relativity simulation of a post-merger waveform~\citep[][black dashed curve]{Dietrich2018, Core2022}. We compute two analytical approximations: Waveform One with an overlap of 0.82 with the numerical-relativity waveform, and Waveform Two with an overlap of of 0.79 with Waveform One. The second waveform is used to test the robustness of the trigger selection algorithm to variations in the frequency structure of the waveform.} 
\end{figure}

In a realistic gravitational-wave search, a template bank would be used to select the optimal match between data and filter. Instead, we select two sets of filter parameters (Table \ref{tab:params}) based on obtaining good overlap (0.82) between our Filter 1 (red, Fig. \ref{fig:waveform}) and the numerical-relativity simulation THC0036~\citep{Dietrich2018, Core2022}.
This numerical-relativity waveform uses the SLy equation of state \citep{SLyEoS} for a $1.35 + 1.35\,\mathrm{M_\odot}$ neutron-star merger (Fig \ref{fig:waveform}, dashed line).
We also use filter 2  (blue, Fig. \ref{fig:waveform}), which has a comparable overlap (0.79) with Filter 1, and is used to study the dependence of our algorithm on the choice of features in the waveform.
For example, Filter 2 maintains a high overlap with Filter 1 while lacking the notch feature at $\simeq 2\,\mathrm{kHz}$. 
While these overlaps are less than what might be considered optimal for a typical matched-filter search, or for parameter-estimation applications, we wish to represent numerical-relativity waveforms as agnostically as possible to study the false-alarm rates irrespective of the underlying waveform. 

\begin{table*}
	\centering
	\caption{Parameters for the waveform used to construct filters in this work. The gravitational wave amplitude $H$ is normalized during the filter whitening step in the \texttt{pycbc.filter.matchedfilter} routine
	}
	\label{tab:params}
	\begin{tabular}{clcc}
		\hline
		Parameter & Description & Value (Filter 1) & Value (Filter 2)\\
		\hline
        $\xi_j$ & Relative weighting of each mode, where $\sum_j \xi_j =1$ & [0.35, 0.56, 0.09] & [0.40, 0.54, 0.06]\\
        $T_j$ & Damping constant of jth mode & [3.0, 1.25, 1.5] ms & [3.5 , 1.9, 0.1] ms\\
        $f_j$ & Frequency of jth mode & [3305, 2540, 1870] Hz & [3250, 2500, 1890] Hz\\
        $\psi_j$ & Relative phase of jth mode & [1.74, 0.68, -0.2] & [1.57, 0.7, 1.57]\\ 
		\hline
	\end{tabular}
\end{table*}

We use a network consisting of the Hanford (H1) and Livingston (L1) detectors. We omit the Virgo detector from this work due to its lower sensitivity. It is trivial to extend the work presented here to include additional detectors, however to afford a significant gain in discovery power it is important that they have comparable sensitivities and glitch rates to the most sensitive detectors.
We use publicly-available O3a data from the Gravitational-Wave Open Science Center~\citep[GWOSC;][]{GWOSC}, and determine how frequently noise processes alone can produce an event trigger that is considered `signal-like'. 
For a given data stream from our detectors, we analyse each second of data and select a trigger using the following procedure:
\begin{enumerate}
\item Identify all triggers that have $\rho_d \ge 4$ in detector $d$ (this process is repeated for each detector in our network).
\item Astrophysical signals must be time-coincident within the light travel time between detectors. We find the loudest trigger in the second detector in a time window [$t_d - \Delta t_L$, $t_d + \Delta t_L$] where $t_L$ is the light travel time between detectors, and $t_d$ is the time at which the trigger with signal-to-noise $\rho_d$ occurs.
\item We calculate the network signal-to-noise ratio $\rho_c$ for each coincident pair of $\rho_d$, where
$$
\rho_\mathrm{c}^2 =\sum_{d=\mathrm{H1,L1}}\rho_d^2.
$$
\item The trigger with the largest $\rho_\mathrm{c}$ is then considered to be a 'signal-like' candidate in that second of data and is stored.
\end{enumerate}
We do not impose a minimum signal-to-noise threshold on the second (non-triggering) detector candidate. Instead, we choose the most significant trigger in the coincidence window. 
We make this choice as we are likely to be dealing with signals that are considered to be sub-threshold and close to the noise level in the detector. 

%
%

We use $16384\,\mathrm{s}$ of public LVK data obtained from the Gravitational Wave Open Science Center\footnote{https://www.gw-openscience.org/}.
We obtain four 4096-second frame files from near the beginning of the third observing run (O3) by the Hanford and Livingston interferometers, beginning at GPS time $1238175744$. 
The strain data is sampled at $16\,\mathrm{kHz}$. We do not down-sample the data, 
however we only analyse calibrated data between $1000$ and $4096\,\mathrm{Hz}$ where the filter waveform is well-defined.
We precondition the data using a highpass filter at $20\,\mathrm{Hz}$.
The median PSD is re-estimated using the Welch method using a 4-second window for each $128\,\mathrm{s}$, non-overlapping block of data.
We remove artefacts at the beginning and end of each 128s block  by windowing the data. These artefacts are introduced by the data pre-conditioning and filter wrap-around, and hence we lose around 4 s of each 128 s data block.
We do not pre-whiten the data or filters as this step is included in the \texttt{pycbc.filter.matchedfilter} routine. 
We do not apply any pre-defined vetos during candidate selection; we defer discussion of the application of conventional LVK vetos to the next section.
In a real search scenario, the non-time coincident background would be collected using the conventional timeslides technique. However, because we only analyze noise data in this work, performing timeslides is only needed if one wishes to build up a larger number of background triggers from which to calculate the false alarm rate. 
To understand the differences between analysis of real and Monte Carlo data, we generate Gaussian noise with an equivalent length and sampling rate generated from the O3a PSD, and perform identical analyses to that described above.

\section{Results \& Discussion}\label{section:results}
\begin{figure}
	 		\includegraphics[width=\columnwidth]{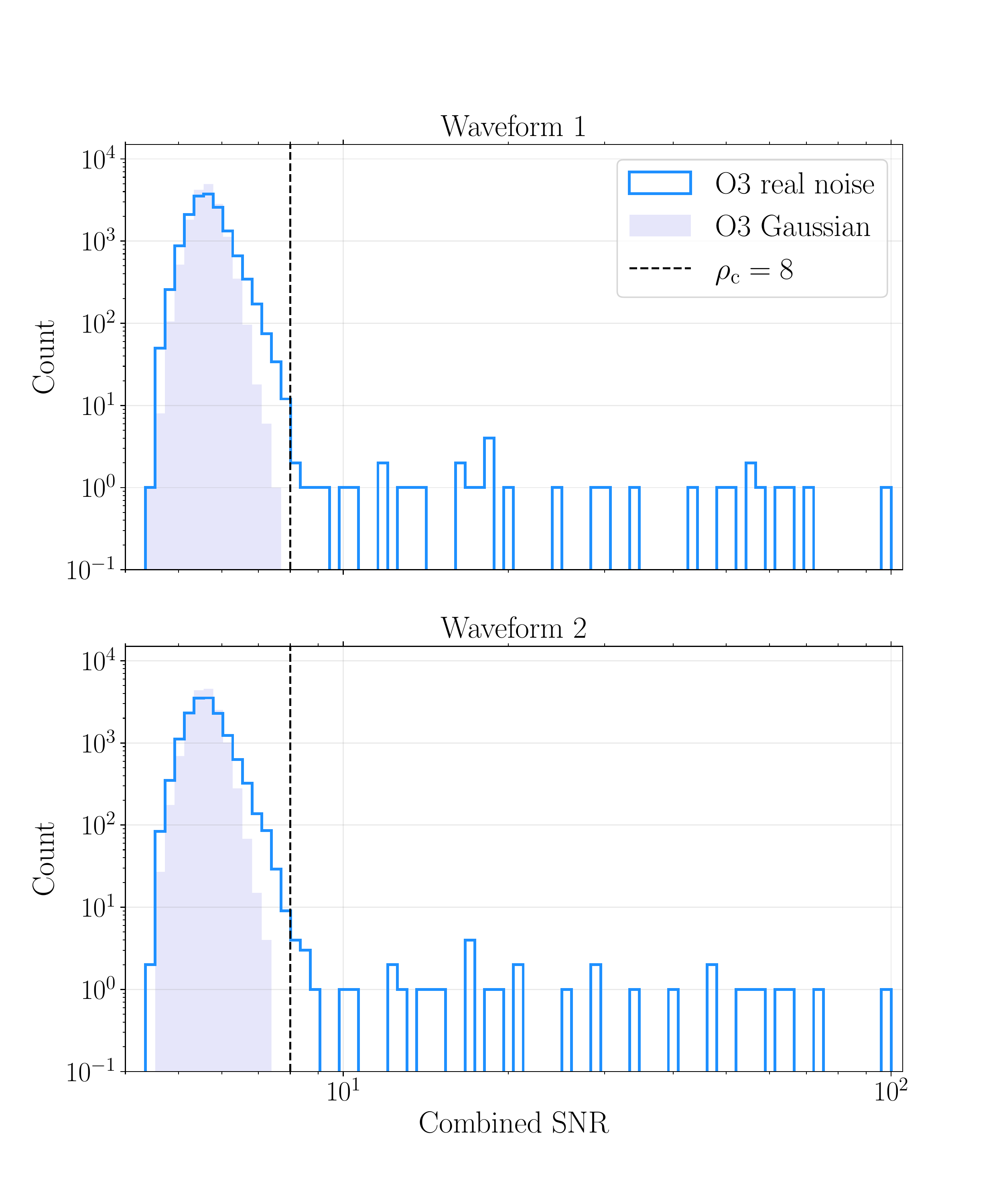} 
	 		\caption{\label{fig:histogram1} Distribution of combined (i.e., network) signal-to-noise ratios (SNRs) from noise triggers identified by the algorithm for Gaussian noise colored by the O3 LIGO Livingston and Hanford PSD (shaded) and real O3 data (unshaded). A long tail of high signal-to-noise ratio noise events is present in the real O3 data, representing departures from stationary Gaussian noise. The distributions are qualitatively (and quantitatively) similar irrespective of the specific waveform filter used (top versus bottom panel), demonstrating the general applicability of the method} 
\end{figure}

To investigate the statistical properties of the noise present at frequencies $>500\,\mathrm{Hz}$, we consider the signal-to-noise distribution of noise triggers identified by our algorithm. Triggers identified from real noise (unshaded histogram, Fig. \ref{fig:histogram1}) with signal-to-noise ratios between 4-6 follow a distribution that is similar to that of Gaussian noise (shaded histogram, Fig. \ref{fig:histogram1}). However at $\rho_\mathrm{c}\gtrsim5$, there is an excess of triggers identified in the real noise data, giving rise to a long tail representing non-stationary and non-Gaussian noise present in the detectors. In general, for each of these triggers, the combined signal-to-noise ratio is dominated by one of the detectors. It is also worthy of note that there is no significant difference in statistical properties between the top and bottom panels of Fig.~\ref{fig:histogram1}, which are the results of the analysis with the two different waveform filters.

We use the histogram of noise triggers in Fig. \ref{fig:histogram1} to visualize how the long tail of triggers may affect the `ranking statistic'. The ranking statistic is used to compute the probability that a putative signal trigger with signal-to-noise ratio $\rho_\mathrm{s}$ arises from this noise distribution alone, and a variety of algorithms exist to compute this statistic (for example, Section D1 of \cite{SPIIRO3} or IIA of \cite{Davies2020}). If there is a long tail of noise triggers a signal trigger will be deemed more likely to have arisen from the underlying noise distribution. The false-alarm probability one then computes from this ranking statistic will be significantly higher, potentially leading to us incorrectly discounting a detection of a real signal as noise. 

\begin{figure}
	 		\includegraphics[width=\columnwidth]{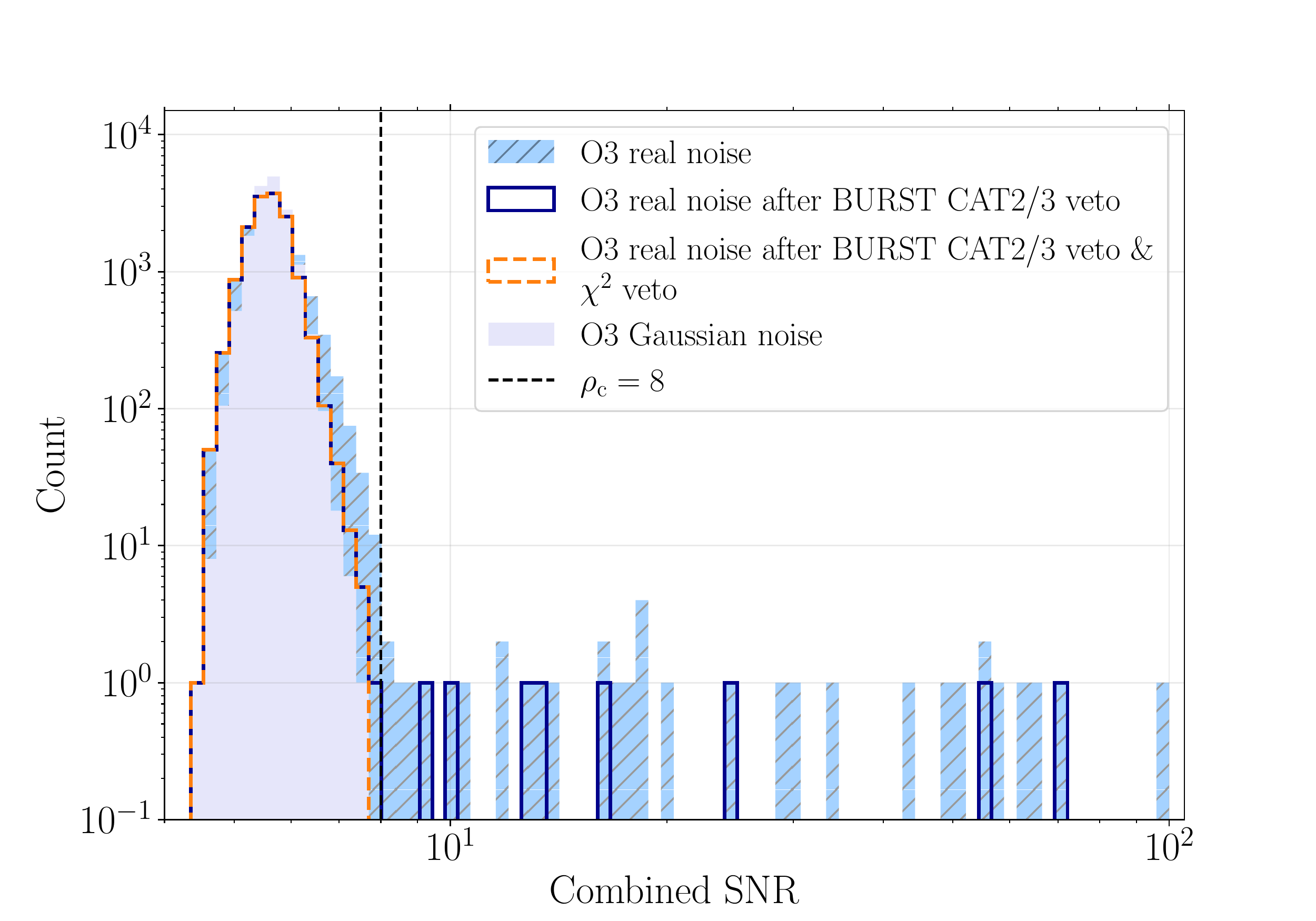} 
	 		\caption{\label{fig:final_hist} Signal-to-noise ratio distribution with triggers removed after applying various vetos. The thatched-blue histogram are the original triggers on O3 real data shown in Fig.~\ref{fig:histogram1}. The dark-blue histogram is after application of the Category 2 and 3 vetos (CAT2/3; see text), and the orange histogram also includes the $\chi^2_r$ veto; this final distribution is in good agreement with the expectation from Gaussian noise.}
\end{figure}

\begin{figure}
	 		\includegraphics[width=\columnwidth]{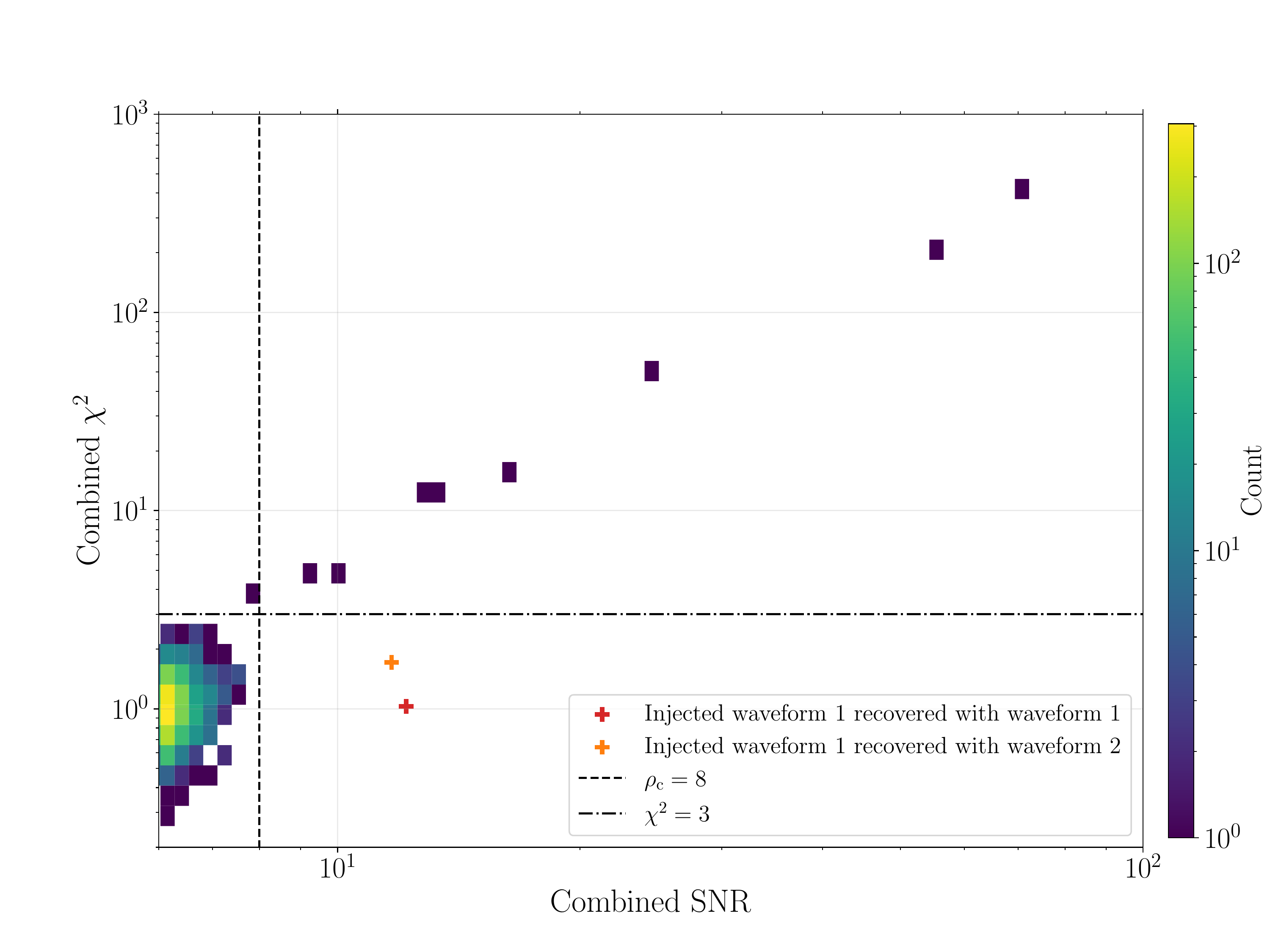} 
	 		\caption{\label{fig:chisqdist} Combined signal-to-noise vs combined $\chi^2$ distribution for triggers with $\rho_\mathrm{c}>6$. If one considers the SNR dimension alone, it is a challenge of disentangling signals from noise using signal-to-noise alone. The 2D histogram demonstrates how using an additional dimension - the $\chi^2$ statistic distribution - allows for both injections to be identified despite the presence of equally or louder noise triggers, and that signals are clearly statistically distinct in this parameter space.}
\end{figure}

The long tail of triggers recovered from the real noise data is undesirable. However they can mostly be removed by applying a standard veto used in short duration searches \citep{Davis2021}.

We apply vetos based on detailed analysis and characterization of the state of the detectors~\citep{Davis2021}.  These vetos, also used in LVK searches for short-duration burst sources, enable us to reject any trigger that occurs at a time when there is a known glitch or data-quality issue in the gravitational-wave strain data. Category 1 (CAT1) vetos indicate times when data should not be used in astrophysical analyses. All data used in our analysis pass the CAT1 veto criterion. A number of times are, however, affected by Category 2 and Category 3 (CAT2 and CAT3) vetos, which indicate noise artefacts due to terrestrial noise. We remove any triggers at times of CAT2 and/or CAT3 vetos. The remaining triggers after application of these vetos are shown as the dark-blue histogram in Fig.~\ref{fig:final_hist}. We see that this veto has removed a number of excess triggers with a combined signal-to-noise ratio $\rho_c\gtrsim5$. Nevertheless, there are a number of loud triggers still present. 

These remaining triggers may still bias the ranking statistic if we rely on $\rho_c$ as our sole statistical measure of the noise properties. Gravitational-wave searches for binary neutron star and black hole inspirals often apply an additional statistical vetos during the process of trigger selection. Although designed to reject non-stationary and non-Gaussian noise events in broadband events, we find that the $\chi^2$ time-frequency discriminator~\citep{Allen2005} can also be effectively employed in our search. For each trigger produced by our algorithm we also compute the corresponding reduced $\chi^2$ value (denoted $\chi^2_r$). We set the number of equal-power frequency bins to ten based on the assumption that for white Gaussian noise $\langle \chi_r^2\rangle \simeq 1$. We calculate the combined $\chi^2_r$ of our triggers as 
$$
\chi^2_{r,c} = \frac{1}{2} \sum_i{\chi^2_{r,i}},
$$ 
where $i$ indexes each detector. In Fig.~\ref{fig:chisqdist} we show our triggers binned in two dimensions: in combined signal-to-noise ratio and combined $\chi^2_r$.

We compare our noise triggers in Fig.~\ref{fig:chisqdist} with a hypothetical gravitational-wave detection. To that end, we inject Waveform One into real detector noise, with injected signal-to-noise ratio $\rho_c = 13.5$, and recover it with each filter shown in Fig. \ref{fig:waveform} in turn (orange and red points in Fig. \ref{fig:chisqdist}). This mimics the detection of a real astrophysical signal. We see that the combined $\chi_r^2$ of these two signals is consistent with that of the noise triggers (i.e., $\chi_r^2\lesssim3$) allowing us to easily distinguish between them and the cluster of noise triggers. Our final veto is therefore defined such that we reject all triggers with $\chi^2_r > 3$.

The dashed-orange histogram in Fig.~\ref{fig:final_hist} shows the final result of applying our $\chi^2_r$ veto in combination with the standard LVK burst vetos. Importantly, this distribution of triggers is extremely close to the idealized Gaussian-noise distribution (light-grey histogram). 

It is worth one word of caution in application of these veto methods to real searches. In reality, the unknown nature of the waveform for post-merger signals may be a potential issue. In our illustrative example, we find that even a $\sim 20$\% mismatch in waveform can increase the value of $\chi^2_r$ by around $50\%$. Therefore, if we do not well-cover our waveform parameter space, applying such a veto may result in missing a real signal. 

With the above caveat in mind, we find that glitches can, and do, contaminate our data even at kHz frequencies. In the portion of data we analyse, we find that Hanford data are subject to CAT2/3 vetos $\sim 1\,\%$ of the time, and Livingston data are subject to the same vetos $\sim 21\,\%$ of the time. It should be noted that the segments chosen may not be representative of the entirety of observing time; for the duration of the O3a observing run, Hanford and Livingston data are subject to the same vetos $\sim 1-2\,\%$ of the individual detector livetimes \citep{Davis2021}. This is a small fraction of data, however it is well known that glitches co-occuring with astrophysical signals have posed challenges to data analysis in the past~\citep[e.g.,][]{GW170817}. In the case of searching for a post-merger signal, most glitches can be effectively vetoed by flagging affected times as not suitable for astrophysical analysis or by utilizing the $\chi^2$ veto, enabling the computation of a more reliable false-alarm rate.

\section{Conclusions}
The observation of gravitational waves from the remnant of a binary neutron star merger has the potential to be a watershed moment in astrophysics, providing unprecedented insight into the hot equation of state of nuclear matter. Such an observation has many technical challenges, not least of which is the fact that the gravitational-wave signal is likely high frequency, short lived, and probably cannot be accurately modelled as with the inspiral phase of these mergers. This raises the issue that such a gravitational-wave signal could be mistaken for one of the many types of short-lived noise artefacts present in gravitational-wave observatories. 

In this work, we show that a series of vetos, including data-quality vetos and the Allen $\chi^2$ veto, can be used to reduce the number of high signal-to-noise ratio false alarms in gravitational-wave data that may mimic a true gravitational-wave signal. To the best of our knowledge, this is the first time the Allen $\chi^2$ statistic has been applied to such short-lived, high-frequency signals, and this work provides a much-needed step towards being able to calculate a false-alarm rate of such signals required to quantify the confidence in a putative detection.

It is premature to calculate a false-alarm rate for a given post-merger gravitational-wave signal. There are no candidate signals in O3 data, and such a false-alarm rate needs to be re-calculated for every observing run (and potentially subsets of observing runs) as the noise statistics are subject to change. Case-in-point is the fact that Livingston data early in the O3a observing run suffered from a 21\% CAT2/CAT3 rate, which averaged to less than 2\% over the entire run. The calculation of a false-alarm rate must therefore utilise enough data when the observatories are in similar states to when the putative detection is made. Following the CAT1/CAT2/CAT3 and Allen $\chi^2$ vetoes, a false-alarm rate using `like' data can easily be calculated alongside signal injections that quantify both false-alarm and false-dismissal rates. We note that this is also subject to change with the advent of potential new and improved data-analysis and processing methods; a point to which we return below.

The first detection of post-merger remnants may actually come from coherently or incoherently stacking multiple sub-threshold signals~\citep[e.g.,][]{Yang2018,Criswell2022}. Studies exploring the mechanics of such detections, as well as astrophysical and equation-of-state inference from such detections, have, to date, only been performed in simulated Gaussian noise. Non-Gaussian and non-stationary artefacts could still play a debilitating role in such analyses, a topic we leave for future work.

It is important to note that a post-merger signal that co-occurs with a glitch may pose a significant challenge to existing parameter-estimation algorithms that assume the signal is present in Gaussian noise only. To mitigate any mis-interpretation of the signal, analysis of such a signal would require simultaneous modelling of the glitch using a method like BayesWave, by incorporating a glitch model into parameter-estimation workflows, or development of new methods that allow for non-Gaussian noise processes \citep[e.g.][]{Ashton2023}. Further development of such would enable more precise measurements of the hot nuclear equation of state, even for the faintest signals. 

We show in this work that the impact of high-frequency noise artefacts in real gravitational-wave data may be somewhat ameliorated with various stages of data vetoes. Of course, this does not fix the problem that gravitational waves from post-merger remnants are incredibly difficult to detect, and likely require another order of magnitude improvement in detector sensitivity in the kHz regime~\cite[e.g.,][and references therein]{Clark2016, abbott17_postmerger,abbott19_postmerger}. Hopefully such improvements will come in the not-too-distant future, with proposed so-called 2.5-generation detectors such as the Neutron Star Extreme Matter Observatory~\citep[NEMO; ][]{NEMO} and 3rd-generation detectors such as the Einstein Telescope~\citep{ET} and Cosmic Explorer~\citep{CE}.

\section*{Acknowledgements}
We acknowledge the rightful owners of the land this research was
conducted on, the Whadjuk Noongar and Bunurong Peoples of the Kulin Nation, and pay our respects to elders past and present.
We are grateful to Eric Thrane and Alex Nitz for helpful suggestions. FHP is supported by a Forrest Research
Foundation Fellowship. This work is supported through Australian Research Council (ARC) Centre of Excellence CE170100004, Discovery Projects DP220101610 and DP230103088, and LIEF Project LE210100002.
This material is based upon work supported by NSF's LIGO Laboratory which is a major facility fully funded by the National Science Foundation. The authors are grateful for computational resources provided by the OzSTAR Australian national facility at Swinburne University of Technology. This research has made use of data or software obtained from the Gravitational Wave Open Science Center (www.gw-openscience.org), a service of LIGO Laboratory, the LIGO Scientific Collaboration, the Virgo Collaboration, and KAGRA. LIGO Laboratory and Advanced LIGO are funded by the United States National Science Foundation (NSF) as well as the Science and Technology Facilities Council (STFC) of the United Kingdom, the Max-Planck-Society (MPS), and the State of Niedersachsen/Germany for support of the construction of Advanced LIGO and construction and operation of the GEO600 detector. Additional support for Advanced LIGO was provided by the Australian Research Council. Virgo is funded, through the European Gravitational Observatory (EGO), by the French Centre National de Recherche Scientifique (CNRS), the Italian Istituto Nazionale di Fisica Nucleare (INFN) and the Dutch Nikhef, with contributions by institutions from Belgium, Germany, Greece, Hungary, Ireland, Japan, Monaco, Poland, Portugal, Spain. KAGRA is supported by Ministry of Education, Culture, Sports, Science and Technology (MEXT), Japan Society for the Promotion of Science (JSPS) in Japan; National Research Foundation (NRF) and Ministry of Science and ICT (MSIT) in Korea; Academia Sinica (AS) and National Science and Technology Council (NSTC) in Taiwan.

\section*{Data Availability}
Code available on reasonable request to authors.



\bibliographystyle{mnras}
\bibliography{neutronstar} 




\bsp	
\label{lastpage}
\end{document}